# Terahertz sensing by beryllium and silicon δ-doped GaAs/AlAs quantum wells

D. Seliuta, J. Kavaliauskas, B. Čechavičius, and S. Balakauskas
Semiconductor Physics Institute, A. Goštauto 11, LT-01108 Vilnius, Lithuania

G. Valušis
Semiconductor Physics Institute, A. Goštauto 11, LT-01108 Vilnius, Lithuania,
and Institute of Materials Science and Applied Research, Vilnius University,
Sauletekio ave. 9, LT-10222, Vilnius, Lithuania

B. Sherliker and M. P. Halsall
School of Electrical and Electronic Engineering, University of Manchester
Manchester, United Kingdom

P. Harrison, M. Lachab, S.P. Khanna, and E.H. Linfield
School of Electronic and Electrical Engineering, University of Leeds,
Leeds LS2 9JT, United Kingdom

Selective sensing of terahertz (THz) radiation by beryllium and silicon δ-doped GaAs/AlAs multiple quantum wells (MQWs) is demonstrated. A sensitivity up to 0.3 V/W within 0.5-4.2 THz in silicon- and up to 1 V/W within 4.2-7.3 THz range in beryllium-doped MQWs at liquid helium temperatures is shown. The built-in electric fields as estimated from the observed Franz-Keldysh oscillations in photoreflectance spectra were found to be located close to the cap and buffer layers of MQWs and vary from 18 kV/cm up to 49 kV/cm depending on the structure design.

The expressed evolution of terahertz (THz) applications and recent progress in THz quantum cascade lasers [1-3] inspires an intensive search for new solutions in design and fabrication of compact receivers for this frequency range. Among the multitude of possible THz sensors, semiconductor nanostructures-based elements deserve a particular interest due to possibility to engineer their properties by design of layer structure and doping profile. As a rule, such devices are compact in dimensions, reliable and are convenient to employ in signal processing. It is worth also noting their advantages in point of view of mass fabrication and integration by usage of semiconductor device technology. A typical example of such an approach can be THz quantum well infrared photo detectors (QWIPs) [4]. Their principle, as in conventional QWIPs [5], relies on bound-to-continuum transitions in a specially-designed (usually GaAs/AlGaAs) structure. However, the principle itself, due to small THz quanta, restricts the operation range to low temperatures; moreover, high absorption in the *Reststrahlen* band of GaAs within the range of 8-9 THz does not allow to make a standard QWIP design suitable for the whole THz range. More delicate approach would be implementation of quantum cascade lasers (QCL)-like structures as THz QWIPs based on a GaAs/AlGaAs superlattice [6]. The idea is based on the electrons trapping in the subsequent active well as soon as they have made their contribution to the photocurrent. It results in a good noise behavior (detectivity of $5 \times 10^7$ cm√Hz/W), but the cost for that is a small responsivity of about 8.6 mA/W. The device operates at 3.5 THz and temperatures up to 50 K.

The other interesting concept is heterojunction interfacial work function internal photoemission (HEIWIP) detectors containing either GaAs [7] or AlGaAs emitters [8]. The latter design allows one to shift the operation frequency to low frequencies, down to 2.3 THz, enabling thus to detect the radiation within broad, 2.3-30 THz, frequency range at low temperatures. Under bias electric field of 2.0 kV/cm, the peak values of responsivity and the specific detectivity in these devices at 9.6 THz and 4.8 K were found to be of 7.3 A/W and $5.3 \times 10^{11}$ Jones, respectively.

Very recently, a new way of utilizing the dopant-assisted intrasubband absorption mechanism in quantum wells (QWs) for normal-incidence THz radiation detection was presented [9]. The proposed concept of the QW intrasubband photodetector (QWISP) is a compact semiconductor heterostructure device compatible with existing GaAs focal-plane array technology and predicts responsivity up to 0.55 A/W close to 3 THz at low temperatures [9].

In this Letter, we extend the family of semiconductor quantum nanostructures-based THz sensors adding a principle of horizontal carrier transport in δ-doped, both of *n*- and *p*-type, GaAs/AlAs MQWs. This approach employs photoionization of bound carriers in the V-shaped potential of the δ-doped impurity states to delocalized states of the QWs (as a rule, it falls into THz range [10]) by the external THz radiation. Since the photoionization energy is potential profile-dependent, QW width can serve as a tuning tool of THz sensing frequency.

Several MQWs designs were prepared on semi-insulating GaAs substrates by molecular beam epitaxy. The AlAs barrier thickness of 5 nm kept for all structures allowed to avoid an overlap of carrier wavefunctions in adjacent wells. The samples were δ-doped at the well centre either with Be acceptor or Si donor atoms. The main growth parameters of the MQWs are given in Table I, sample's schematic view and design details can be found in Fig. 2 caption.

TABLE I: Characteristics of the studied MQWs: repeated periods, quantum well width $L_w$ and the δ-doping density

| Samples | Periods | $L_w$ (nm) | doping type and density, cm$^{-2}$ |
|---------|---------|------------|-------------------------------------|
| L151    | 60      | 15         | *p*-type, $5 \times 10^{12}$        |
| L152    | 60      | 15         | *p*-type, $5 \times 10^{11}$        |
| 1392    | 40      | 20         | *p*-type, $2.5 \times 10^{12}$      |
| 1794    | 200     | 10         | *p*-type, $5 \times 10^{10}$        |
| 1795    | 400     | 3          | *p*-type, $2 \times 10^{10}$        |
| L78     | 40      | 15         | *n*-type, $1.4 \times 10^{11}$      |
| L79     | 40      | 15         | *n*-type, $4 \times 10^{11}$        |

The MQWs were studied by measuring photocurrent induced by an external THz radiation. Two types of THz sources – either a free electron (FELIX, the Netherlands) or an optically-pumped molecular THz laser – were employed. Before processing, the as-grown structures were characterized by photoreflectance (PR) spectroscopy exciting the structures either by a He-Ne (632.8 nm) laser or LED (470 nm) light beams.

The aim of characterization of MQWs by the PR spectroscopy was checking of presence and a value of built-in electric fields. The experimental results are given in Fig. 1. As one can see, in moderately and highly doped samples the spectra are dominated by a bulk-like oscillating signal arising above of the GaAs fundamental gap $E_g$ (1.424 eV at 300 K). These features in PR spectra are associated with Franz-Keldysh oscillations (FKOs) indicating the existence of built-in electric fields in the structures. By analysing the FKOs data, (i. e. using the relationship between the energy position $E_m$ of FKOs extrema and the extremum index *m* and plotting the dependence of the quantity $(4/3\pi)(E_m-E_g)^{3/2}$ as a function of index *m* one can extract the value of the built-in electric field strength. As it is seen, the values vary within 18-49kV/cm range depending on the structure design [11].

Additionally, our results on FKOs and surface photovoltage spectroscopy [12] (not given here) allowed one to determine that the built-in electric fields are located mainly in the cap and buffer layers of GaAs. Therefore, they do not ionize the bound states in MQWs region and hence should have no essential effect for the THz detection in the proposed scheme.

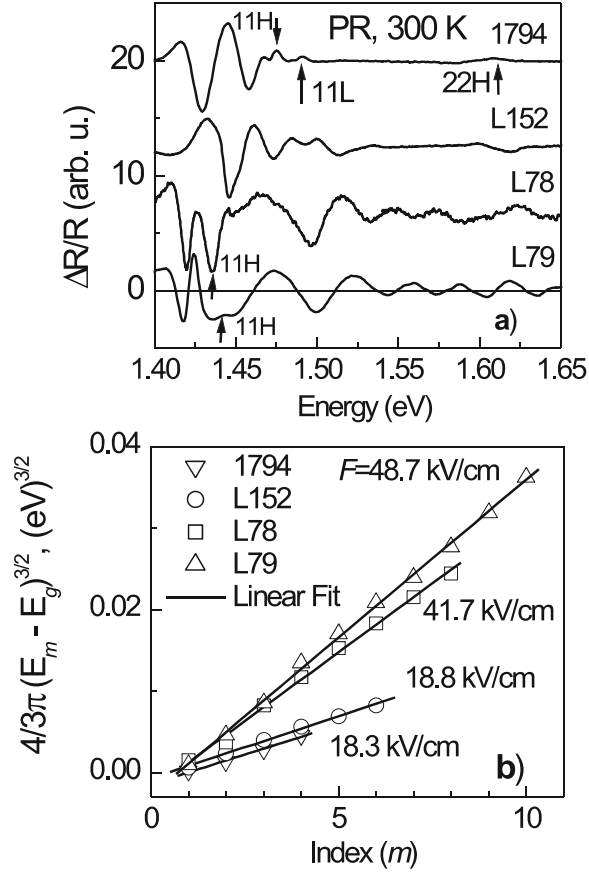

FIG. 1. The PR spectra (a) exhibiting FKOs at 300 K. Note that the oscillations are superimposed with excitonic transitions. Plot (b) serves for estimates of the built-in electric field using the FKOs data and a procedure described in Ref. 11.

The THz spectroscopic data are presented in top plot of Fig. 2. It can be seen that there is a fairly narrow band absorption streching from 50-57 μm (4.9-5.6 THz) and centered around 55 μm (5.4 THz). This correlates well with earlier Fourier transform and balanced pump-probe spectroscopy measurements which revealed a strong absorption at 55 μm due to intra-acceptor excitation of the bound hole from the 1s to the 2p state (the D-line) [13]. Therefore, it is obvious that the photocurrent measured here is of the same origin.

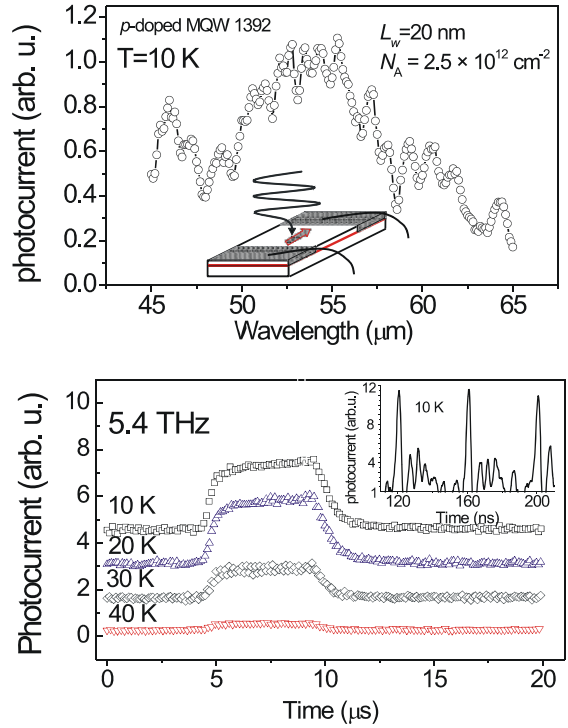

FIG. 2. (Color online) The FELIX data. Top plot - THz photocurrent spectra and schematic view (inset) of the sample for THz photocurrent experiments: two-electrode of 1.5×1 mm dimensions and distance in between of 1.5 mm were prepared as mezas of 3 μm hight using conventional optical lithograpghy and wet etching. Ohmic contacs were fabricated by rapid annealing procedure of Au/Ge/Ni in *n*-type and Au/Cr alloys in *p*-type samples, respectively. Bottom plot -- THz current transients at 5.4 THz (55 μm) at various lattice temperatures. The inset depicts a close-up view of a small region of the temporal photocurrent response.

The bottom plot of Fig. 2 depicts the photocurrent as a function of time across one of the free-electron laser's macropulses [14] recorded at various temperatures. As one can see, the signal decreases with temperature and at 40 K becomes hardly resolved. The inset shows a close-up view of a small region of the temporal response of the photocurrent. It can be seen that the variation in the data is actually series of very narrow peaks detected at regular repetition rate – the structure responds to each 10 ps micropulse of the free electron laser with the width of each signal peak around 4 ns. We attribute it to the capture time of the photoionized free holes back onto the ionized impurities as it was evidenced in persistent photoconductivity experiments [15].

Results of the experimental investigation of bound-to-continuum transitions in *p*-type doped MQWs, using optically-pumped molecular THz laser are shown in top plot of Fig. 3. The signal is attributed to the photothermal ionization of the Be acceptors by the external THz radiation. It is worth noting three distinctive features in *p*-type doped MQWs: *first*, the photocurrent exhibits threshold-like behavior as expected for photoionization process; *second*, the signal appears far below the ionization energy (under 4.2 THz, while the Be binding energy amounts, e. g. 32 meV (7.8 THz) for 15 nm QW [10]) – it is probably related to the contribution of excited acceptor states $2P_{3/2}$ and $2P_{5/2}$ having transition energies several meV lower than the binding energy[16]; *third*, the THz photocurrent signal depends nonmonotonically on the doping density as is depicted in the inset: up to about $10^{12}$ cm$^{-2}$ it increases (because of an increased absorption) while above this value it drops down due to the approach of the Mott transition – this assumption is also supported by photoluminescence and modulation spectroscopy data [18].

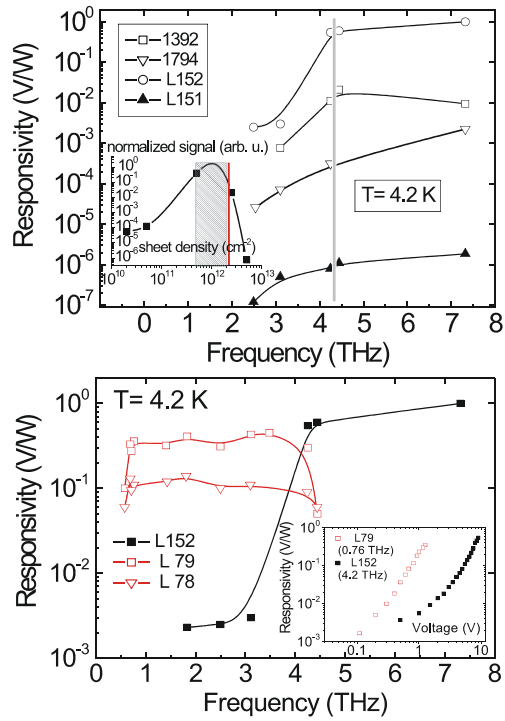

FIG. 3. Optically-pumped THz laser data. Top plot: Responsivity as a function of frequency in various *p*-type doped MQWs. Grey line – frequency under which THz responsivity is normalized. Inset illustrates nonmonotonical behavior of the normalized (to 40 MQWs periods) THz photocurrent with the hole sheet density. Red line – theoretical estimate of the Mott transition [17], shaded area indicates the range of doping density for optimal THz detection. Bottom plot: Comparison of responsivity as a function of frequency in *n*- and *p*- type doped MQWs. Inset -- responsivity vs. voltage in L79 and L152 samples.

To cover the "red side" of the THz spectrum, we have used silicon doped MQWs [19]. Results of the experimental study are shown in bottom plot of Fig. 3. Similarly to the case of $p$-type doped MQWs, the photocurrent signal, due to photothermal ionization of Si donors by the incident radiation, appears around 0.6~THz (it is below even the bulk binding energy of 5.8 meV (1.41 THz) perhaps because of excitation of intradonor transitions) and manifests *plateau* within the range of 0.6–4 THz. The maximal responsivity amounts to 0.32 V/W.

A noise equivalent power measured at 1 kHz using the conventional lock-in technique was found to be in the range of 5 nW/√Hz at 4.2 K.

In conclusion, we have demonstrated that horizontal carrier transport below 40 K can successfully be employed for the THz detection in Be and Si δ-doped GaAs/AlAs MQWs. The sensitivity up to 0.3 V/W within 0.5–4.2 THz in Si- and up to 1 V/W within 4.2-7.3 THz range in Be-doped MQWs at 4.2 K is demonstrated. The built-in electric fields – from 18 kV/cm up to 49 kV/cm, depending on the structure design - estimated from the observed FKOs oscillations in the PR spectra were found to be located close to the cap and buffer layers of MQWs and do not ionize the bound states in the MQWs region.

We are very grateful for Matthew Steer (Sheffield) for providing samples 1392, 1794 and 1795 as well as Jonathan Philips for the kind assistance during FELIX experiments. The Vilnius group acknowledges the support from the Lithuanian State Science and Studies Foundation (Project No. C-07004) and the EOARD (contract No. FA8655-06-1-3007).